\documentclass[aps,prmaterials,twocolumn,superscriptaddress,amsmath,amssymb,floatfix,longbibliography]{revtex4-2}

\usepackage[T1]{fontenc}
\usepackage{graphicx}
\usepackage{bm}
\usepackage{color}
\usepackage{physics}
\usepackage[colorlinks=true,citecolor=blue,linkcolor=blue,urlcolor=blue,filecolor=blue]{hyperref}
\usepackage{booktabs}

\begin{document}

\title{Evidence of Haldane-Chain Physics in an Fe--Dehydroindigo Coordination Polymer}

\author{Ritam Chakraborty}
\email{ritamchakraborty454@gmail.com}
\affiliation{Theoretical Sciences Unit, Jawaharlal Nehru Centre for Advanced Scientific Research (JNCASR), Jakkur, Bangalore 560064, India}

\author{T.~V.~Ramakrishnan}
\email{tvrama2002@yahoo.co.in}
\affiliation{Theoretical Sciences Unit, Jawaharlal Nehru Centre for Advanced Scientific Research (JNCASR), Jakkur, Bangalore 560064, India}
\affiliation{Department of Physics, Indian Institute of Science, Bangalore 560012, India}

\begin{abstract}
We investigate whether an Fe--dehydroindigo spin-crossover coordination polymer can support Haldane-chain physics in its intrinsic, substrate-free limit. Spin-orbit-coupled density-functional-theory calculations and magnetic energy mapping yield a nearly isotropic antiferromagnetic coupling, together with much weaker single-ion anisotropy. Using these parameters, exact diagonalization and density-matrix renormalization group calculations give a nondegenerate ground state separated from the lowest triplet-derived excitations by a gap of about $13$--$14$~meV, on the scale expected for the spin-1 Haldane gap. The anisotropy produces only a small splitting of the low-energy modes. The dynamical spin structure factor places the lowest spectral weight near the antiferromagnetic wave vector $q=\pi$. These results identify the isolated cis-dehydroindigo chain as a material-specific coordination-polymer platform whose microscopic interactions place it in a gapped regime consistent with Haldane-chain physics.
\end{abstract}

\maketitle

\section{Introduction}
\label{sec:intro}

One-dimensional antiferromagnetic spin chains provide a canonical example of the effect of quantum fluctuations on a many-body ground state. Haldane's conjecture~\cite{haldane1983nonlinear} established a qualitative distinction between half-integer and integer-spin Heisenberg chains. For the isotropic spin-1 chain, the ground state is nondegenerate and separated from the lowest triplet excitation by a finite gap, $\Delta \simeq 0.4105J$~\cite{white1993numerical}. The corresponding Haldane phase is now understood as a symmetry-protected topological phase, with nonlocal string order and fractionalized spin-$1/2$ edge degrees of freedom for open chains~\cite{Pollmann2012}. Early neutron-scattering studies of quasi-one-dimensional spin-1 compounds, including CsNiCl$_3$ and NENP, established the characteristic gapped excitation spectrum~\cite{Buyers1986,renard1987presumption}.

The search for material realizations has since extended to molecular and atomically designed spin chains. First-principles exchange parameters have been used to assess molecular systems such as chromium phthalocyanine~\cite{Wu2013}. More recently, on-surface-synthesized triangulene chains provided an atomic-scale realization in which gapped bulk excitations and fractionalized edge states were observed by scanning tunnelling spectroscopy~\cite{Mishra2021}, while electronic-structure and many-body calculations have explored related Haldane and dimerized regimes in spin-1 nanographene chains~\cite{Henriques2025}. These advances make the materials question more specific: can the microscopic interactions of a chemically distinct coordination polymer place its spin-1 backbone in the same gapped regime without assuming the parameters of an ideal model?

Spin-crossover coordination polymers provide a distinct setting for this question because the local spin state, exchange interaction, and magnetic anisotropy all arise from the metal--ligand coordination environment. Model calculations have shown that one-dimensional spin-crossover chains can host several competing phases, including a Haldane phase for antiferromagnetic exchange and sufficiently weak anisotropy~\cite{Powell2026}. What remains material dependent is whether this hierarchy of interactions occurs in a specific coordination polymer.

\begin{figure*}[t]
    \centering
    \begin{minipage}[b]{0.425\textwidth}
        \centering
        \includegraphics[width=\linewidth]{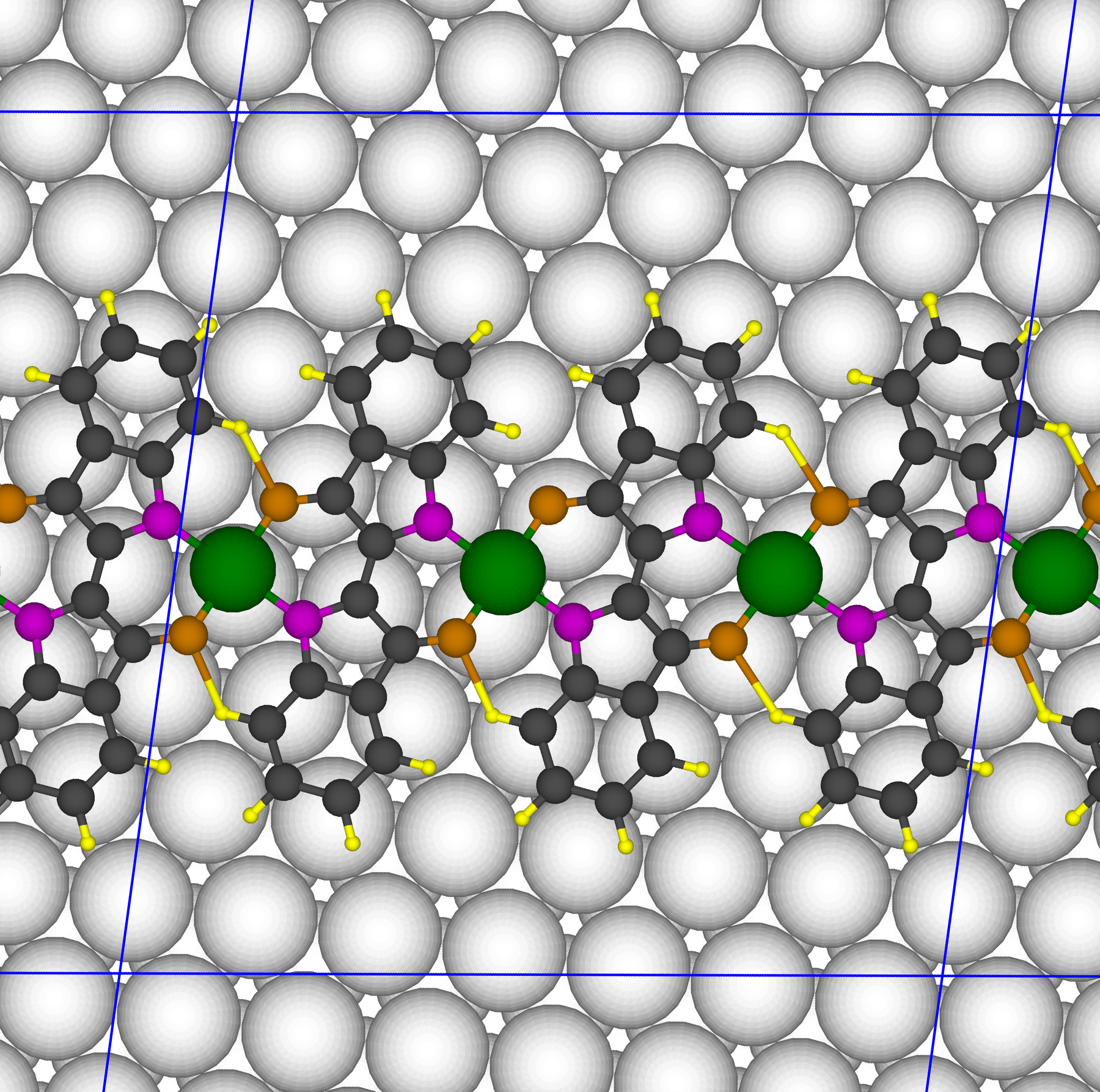}\\[-1mm]
        (a) \textit{trans} CP
    \end{minipage}
    \hfill
    \begin{minipage}[b]{0.42\textwidth}
        \centering
        \includegraphics[width=\linewidth]{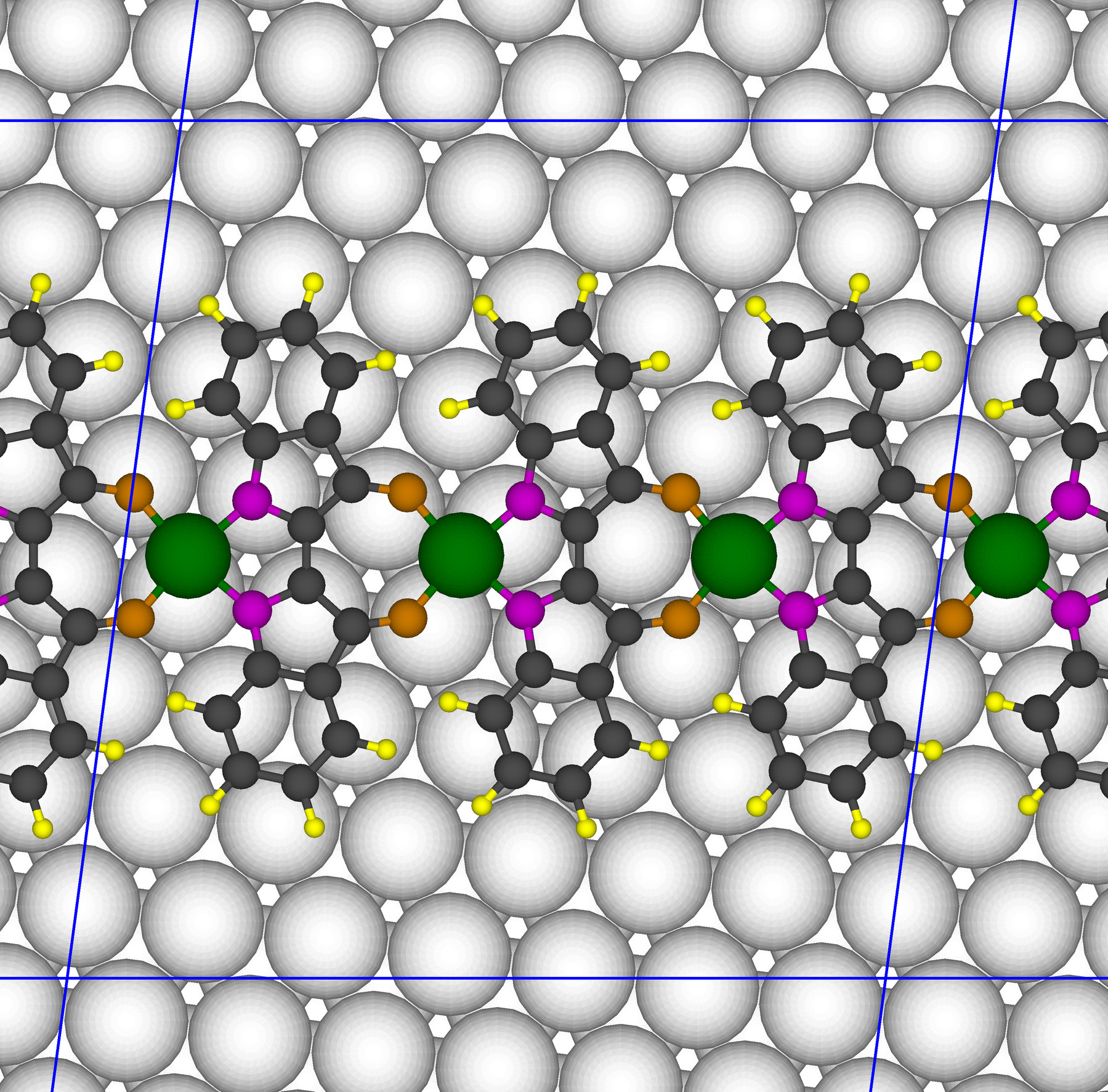}\\[-1mm]
        (b) \textit{cis} CP
    \end{minipage}
    \caption{Surface unit cells used for the \textit{trans} and
    \textit{cis} coordination polymers on Ag(111). Each unit cell
    contains three dehydrogenated indigo molecules and three Fe atoms.
    The structural change modifies the first coordination shell of Fe:
    the \textit{trans} chain is N,O-chelated, whereas the \textit{cis}
    chain contains (N,N)- and (O,O)-chelated Fe environments. Black, yellow,
    pink, orange, green, and white spheres denote C, H, N, O, Fe, and Ag,
    respectively; blue lines mark the surface unit-cell boundary.}
    \label{fig:CP-on-Ag111}
\end{figure*}

Here we address this question for an Fe--dehydroindigo coordination polymer derived from one-dimensional structures synthesized on Ag(111) \cite{xu2024}, shown in Fig.~\ref{fig:CP-on-Ag111}. The spin-crossover behavior of these Fe--indigo chains has recently been studied by first-principles calculations~\cite{sco_paper}. That work identifies the \textit{cis} polymer (shown in Fig.~\ref{fig:CP-on-Ag111}(b)) as the lower-energy Ag(111) structure in the relevant interaction regime and, importantly, finds only uniform (pure) spin configurations for the \textit{cis} chain, in contrast to the mixed-spin patterns possible for the \textit{trans} chain. This makes the \textit{cis} geometry a natural starting point for a uniform spin-chain description: in the $S=1$ interaction regime considered here, all Fe centers are equivalent spin-1 sites. We therefore focus on the isolated periodic \textit{cis}-dehydroindigo chain (\ref{fig:CP-on-Ag111}(b)) as a controlled reference system. This separates the intrinsic magnetic interactions of the polymer backbone from substrate-induced screening, hybridization, and interfacial symmetry breaking. The resulting model is therefore an intrinsic limit of the experimentally realized coordination polymer rather than a complete description of the Ag-supported system.

We determine the exchange and single-ion anisotropy tensors from spin-orbit-coupled density-functional-theory (DFT) calculations using magnetic energy mapping~\cite{XiangWhangbo2013,Udvardi2003} over 42 configurations. The resulting interaction hierarchy is then reduced to an effective spin-1 quantum Hamiltonian and studied by exact diagonalization (ED) and density-matrix renormalization group (DMRG) calculations. We finally calculate the zero-temperature dynamical spin structure factor to connect the low-energy spectrum to a momentum-resolved magnetic response. Because explicit symmetry-protected-topological diagnostics such as string order, the entanglement spectrum, or open-chain edge states are not evaluated here, we use the more conservative description ``Haldane-chain physics'' or ``Haldane-like regime'' for the material-specific results.

\section{Computational framework}
\label{sec:methods}

DFT calculations were performed with the Vienna \textit{Ab initio} Simulation Package (VASP)~\cite{Kresse1996,Kresse1999}. The interaction between the ionic core and valence electrons was described using the projector augmented-wave method~\cite{Blochl1994}. Exchange and correlation were treated
with the nonlocal vdW-DF2-B86R functional \cite{hamada2014van, lee2010higher}. Strong correlations in the Fe $3d$ shell were treated within the Dudarev DFT+$U$ scheme~\cite{Dudarev1998}, with $U=2$~eV and $J=1$~eV, corresponding to $U_{\mathrm{eff}}=1$~eV.

The system under consideration was relaxed in a collinear spin-polarized calculation using a plane-wave cutoff of 550~eV, an electronic convergence threshold of $10^{-7}$~eV, a force threshold of $0.01$~eV/\AA, and a $\Gamma$-centered $4\times1\times1$ $k$-point mesh. Spin-orbit-coupled noncollinear calculations were then performed using the fully relativistic VASP implementation~\cite{Hobbs2000}. A $\Gamma$-centered $80\times1\times1$ mesh was used for the magnetic energy mapping~\cite{XiangWhangbo2013}. The magnetization axis was sampled along 21 spatial directions, with both ferromagnetic (FM) and antiferromagnetic (AFM) configurations, giving 42 total energies. Details of the directions and the validation procedure are given in the Supplemental Material.

The system was subsequently modelled by a spin-only Hamiltonian. The spin-only reduction is supported by the strong-coupling hierarchy: $U_{\mathrm{eff}}=1$~eV and the estimated hopping $t\simeq40$~meV give $U_{\mathrm{eff}}/t\simeq25$, so low-energy charge fluctuations are strongly suppressed. This is the same physical rationale used in a recent organic Haldane-chain construction with $U/t>100$~\cite{Anindya2026}. For the present multiorbital Fe--ligand manifold, we use this ratio only as a scale-separation check; the exchange parameters are obtained directly from total-energy mapping. An inorganic $S=1$ GGA+$U$--to--Heisenberg precedent is discussed in the Supplemental Material.

The DFT-derived spin model was solved using ED with QuSpin~\cite{Weinberg2017,Weinberg2019} for periodic spin-1 chains up to $N=16$. DMRG calculations using TeNPy~\cite{hauschild2024tenpy} were carried out for chains up to $N=100$ as a longer-chain benchmark of the lowest gap. For the anisotropic Hamiltonian, the DMRG calculation conserves total $S^z$ by retaining only the $U(1)$-symmetric part of the single-ion anisotropy, $D_\perp=(D_{xx}+D_{yy})/2$ together with $D_{zz}$; the residual symmetry-breaking term $(D_{xx}-D_{yy})/2\simeq0.032$~meV is three orders of magnitude smaller than $J_{\mathrm H}$ and is not expected to shift the reported gap values within their quoted precision. The full orthorhombic anisotropy, including this term, is retained without approximation in the ED calculation. The zero-temperature dynamical spin structure factor was evaluated for an $N=12$ periodic chain using the Lanczos continued-fraction method.

\section{DFT-derived magnetic Hamiltonian}
\label{sec:tensors}

The magnetic energies were mapped onto the minimal two-Fe repeat unit of the isolated chain. With two equivalent nearest-neighbor bonds per two-site periodic cell, the classical spin Hamiltonian is written as
\begin{equation}
\mathcal{H}_{\mathrm{spin}}
=
-2\sum_{\alpha,\beta}J_{\alpha\beta}S_1^\alpha S_2^\beta
+
\sum_{\alpha,\beta}D_{\alpha\beta}
\left(S_1^\alpha S_1^\beta+S_2^\alpha S_2^\beta\right),
\label{eq:spinham}
\end{equation}
where $\alpha,\beta\in\{x,y,z\}$. The factor of two in the exchange term accounts for the two equivalent nearest-neighbor bonds represented by the periodic two-site cell. The symmetric single-ion anisotropy tensor was constrained to be traceless, since its isotropic part contributes only a constant proportional to $S(S+1)$.

The fitted symmetric exchange tensor, in meV, is
\begin{equation}
\mathbf{J}=
\begin{pmatrix}
-16.001 & -0.024 & 0.020\\
-0.024 & -16.124 & -0.021\\
0.020 & -0.021 & -16.185
\end{pmatrix},
\label{eq:Jtensor}
\end{equation}
and the corresponding traceless single-ion anisotropy tensor is
\begin{equation}
\mathbf{D}=
\begin{pmatrix}
-0.152 & -0.006 & 0.055\\
-0.006 & -0.089 & 0.040\\
0.055 & 0.040 & 0.241
\end{pmatrix}.
\label{eq:Dtensor}
\end{equation}

The exchange tensor is nearly isotropic. Its diagonal average is
\begin{equation}
J_{\mathrm{iso}}
=
\frac{J_{xx}+J_{yy}+J_{zz}}{3}
=
-16.103~\mathrm{meV}.
\end{equation}
With the convention of Eq.~\eqref{eq:spinham}, this corresponds to the standard antiferromagnetic Heisenberg coupling
\begin{equation}
J_{\mathrm H}=-2J_{\mathrm{iso}}=32.206~\mathrm{meV}.
\label{eq:JH}
\end{equation}
The off-diagonal exchange terms are at the level of a few $10^{-2}$~meV. The dominant isotropic component, $J_{\mathrm{iso}}\approx-16.1$~meV, is consistent with antiferromagnetic superexchange transmitted through the $\pi$-conjugated dehydroindigo backbone bridging adjacent Fe centers, rather than through direct Fe--Fe orbital overlap. The near-degeneracy of the diagonal exchange components and the smallness of the off-diagonal terms follow the usual hierarchy expected for a superexchange-dominated antiferromagnet: to leading order the coupling is set by spin-independent kinetic exchange, with spin-orbit coupling entering only as a weak relativistic correction that generates the anisotropic exchange and single-ion anisotropy terms. The single-ion anisotropy is correspondingly small compared with $J_{\mathrm H}$; microscopically, it arises from second-order spin-orbit coupling acting within the low-symmetry Fe ligand field, with its sign and magnitude controlled by the splitting of the Fe $3d$ orbitals -- a decomposition beyond the scope of the present energy-mapping analysis. In the Cartesian frame used here (with $x$ along the chain axis), $D_{zz}>0$ identifies $z$ as the hard direction relative to the $xy$ plane, while $D_{xx}<D_{yy}$ gives a weaker easy-axis-like anisotropy within that plane; associating $y$ and $z$ with specific directions in the relaxed molecular geometry would require examining the corresponding Kohn--Sham orbitals, which we do not attempt here. The classical energy difference between the $x$ and $z$ directions from the diagonal terms is $D_{zz}-D_{xx}=0.393$~meV.

The 42-state mapping was tested by grouped leave-one-direction-out cross-validation, in which the FM and AFM states belonging to the same spatial direction were kept together. The reported validation gives a mean absolute error of 0.13~meV and $R^2=0.9999$, while the fitted tensor elements vary by less than 0.04~meV for $\mathbf J$ and 0.02~meV for $\mathbf D$ across the validation folds. These results show that the bilinear exchange-plus-SIA model describes the sampled magnetic configurations accurately. They do not, by themselves, exclude interactions that are not resolved by this configuration set. The full configuration list and validation details are provided in the Supplemental Material.

For the many-body calculations, the exchange is represented by the isotropic coupling $J_{\mathrm H}$ and the diagonal single-ion terms are retained. The off-diagonal tensor components, all below 0.06~meV, are neglected at this stage. The quantum Hamiltonian is therefore
\begin{equation}
\begin{split}
\hat{H}={}&
J_{\mathrm H}\sum_{i=1}^{N}\hat{\mathbf S}_i\cdot\hat{\mathbf S}_{i+1}\\
&+
\sum_{i=1}^{N}
\left[
D_{xx}(\hat S_i^x)^2+
D_{yy}(\hat S_i^y)^2+
D_{zz}(\hat S_i^z)^2
\right],
\end{split}
\label{eq:ed_hamiltonian}
\end{equation}
with periodic boundary conditions and
\begin{equation}
\begin{split}
D_{xx}&=-0.152~\mathrm{meV},\quad D_{yy}=-0.089~\mathrm{meV},\\
D_{zz}&=0.241~\mathrm{meV}.
\end{split}
\end{equation}
The ratio of the anisotropy scale to the exchange scale is of order $10^{-2}$, placing the model close to the isotropic spin-1 antiferromagnetic Heisenberg chain.

\section{Low-energy many-body spectrum}
\label{sec:manybody}

\begin{figure}[t]
\centering
\includegraphics[width=\columnwidth]{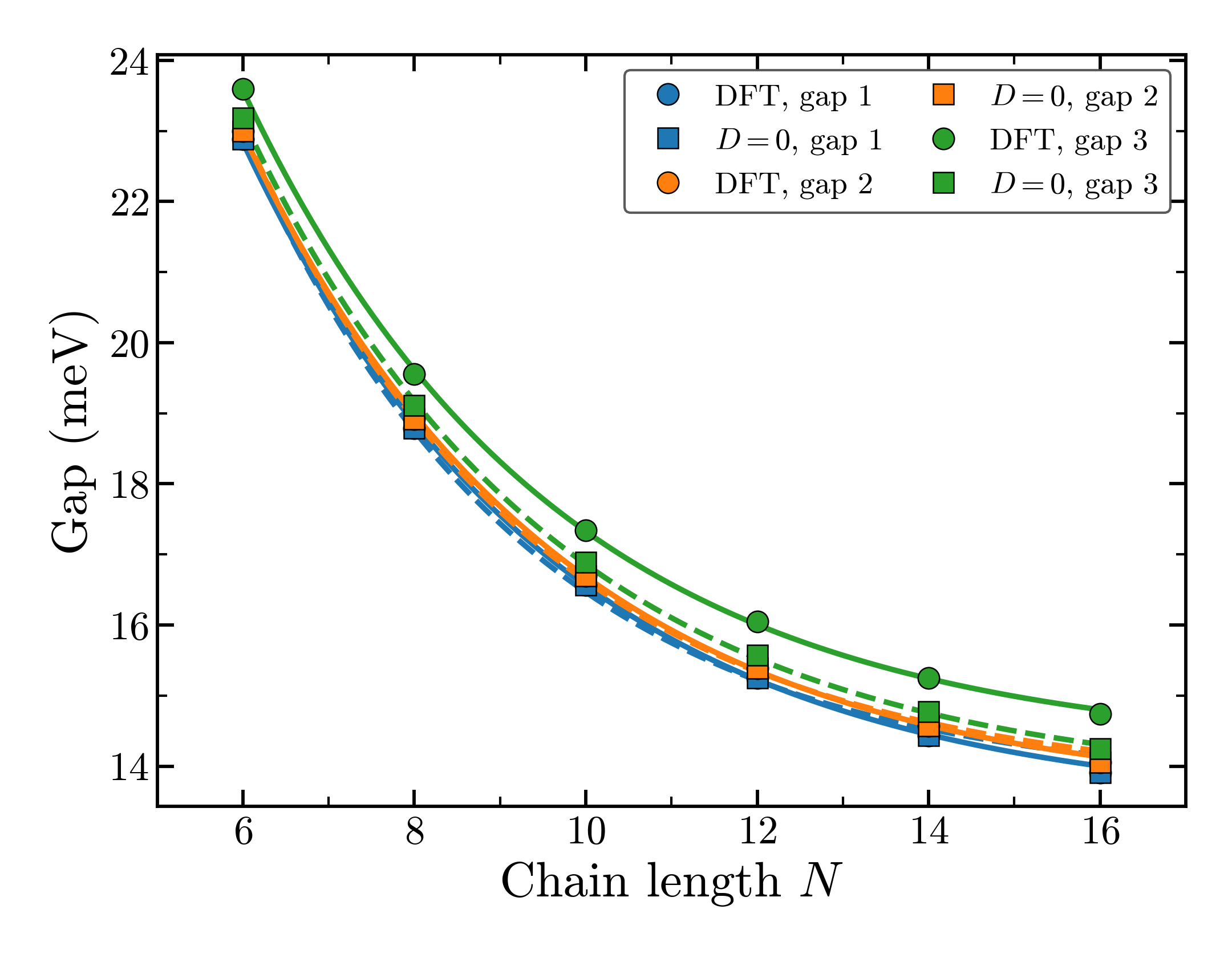}
\caption{Finite-size scaling of the three lowest excitation gaps of the periodic spin-1 chain from exact diagonalization. Circles and solid lines show the model with the DFT-derived diagonal single-ion anisotropy; squares and dashed lines show the isotropic ($D_{\alpha\alpha}=0$) result. Exponential fits give $\Delta_{D=0}(\infty)=13.695$~meV in the isotropic limit and $\Delta_1(\infty)=13.382$~meV, $\Delta_2(\infty)=13.516$~meV, and $\Delta_3(\infty)=14.196$~meV with the anisotropy included.}
\label{fig:ED_results}
\end{figure}

We first examine the low-energy spectrum by ED for periodic chains with $N=6$--16. Figure~\ref{fig:ED_results} shows the three lowest excitation gaps with and without the DFT-derived diagonal SIA. In the isotropic limit, the lowest triplet is degenerate and the exponential finite-size fit gives
\begin{equation}
\Delta_{D=0}(\infty)=13.695~\mathrm{meV}.
\end{equation}
For comparison, the thermodynamic Haldane gap of the isotropic spin-1 Heisenberg chain is $\Delta/J=0.41050(2)$~\cite{white1993numerical}, which at the extracted exchange scale gives
\begin{equation}
\Delta_{\mathrm H}\simeq0.4105J_{\mathrm H}\simeq13.22~\mathrm{meV}.
\end{equation}
The small difference is consistent with the limited chain lengths entering the ED extrapolation.

With the DFT-derived diagonal SIA included, the lowest triplet-derived excitation is weakly split. Using
\begin{equation}
\Delta_a(N)
=
\Delta_a(\infty)+A_a\exp\left(-\frac{N}{\lambda_a}\right),
\label{eq:exp_fit}
\end{equation}
we obtain
\begin{equation}
\begin{split}
\Delta_1(\infty)&=13.382~\mathrm{meV},\\
\Delta_2(\infty)&=13.516~\mathrm{meV},\\
\Delta_3(\infty)&=14.196~\mathrm{meV}.
\end{split}
\end{equation}
Here $\lambda_a$ is only the decay scale of the finite-size fit and should not be identified with the bulk spin--spin correlation length. The higher-energy branch is consistent with the larger cost of out-of-plane fluctuations for $D_{zz}>0$. An alternative inverse-size fit, included in the Supplemental Material, illustrates the systematic uncertainty associated with extrapolating the short ED chains; the main conclusion used here is the persistence of a finite gap on the $J_{\mathrm H}$ scale.

A complementary DMRG calculation was performed for longer chains. In the present implementation, the lowest sector-resolved gap was evaluated as
\begin{equation}
\Delta_{\mathrm{DMRG}}(N)
=
E_0(S^z_{\mathrm{tot}}=1)
-
E_0(S^z_{\mathrm{tot}}=0).
\label{eq:dmrg_def}
\end{equation}
For the isotropic model this directly gives the lowest triplet--ground-state gap. For the anisotropic model, $S^z_{\mathrm{tot}}$ is a good quantum number only for the $U(1)$-symmetric truncation of the SIA described in Sec.~\ref{sec:methods}; $\Delta_{\mathrm{DMRG}}(N)$ is evaluated for this restricted Hamiltonian and used here as a numerical benchmark of the lowest gap scale, while the exact orthorhombic anisotropy and its splitting of the triplet are treated without approximation by ED.

\begin{figure}[t]
\centering
\includegraphics[width=\columnwidth]{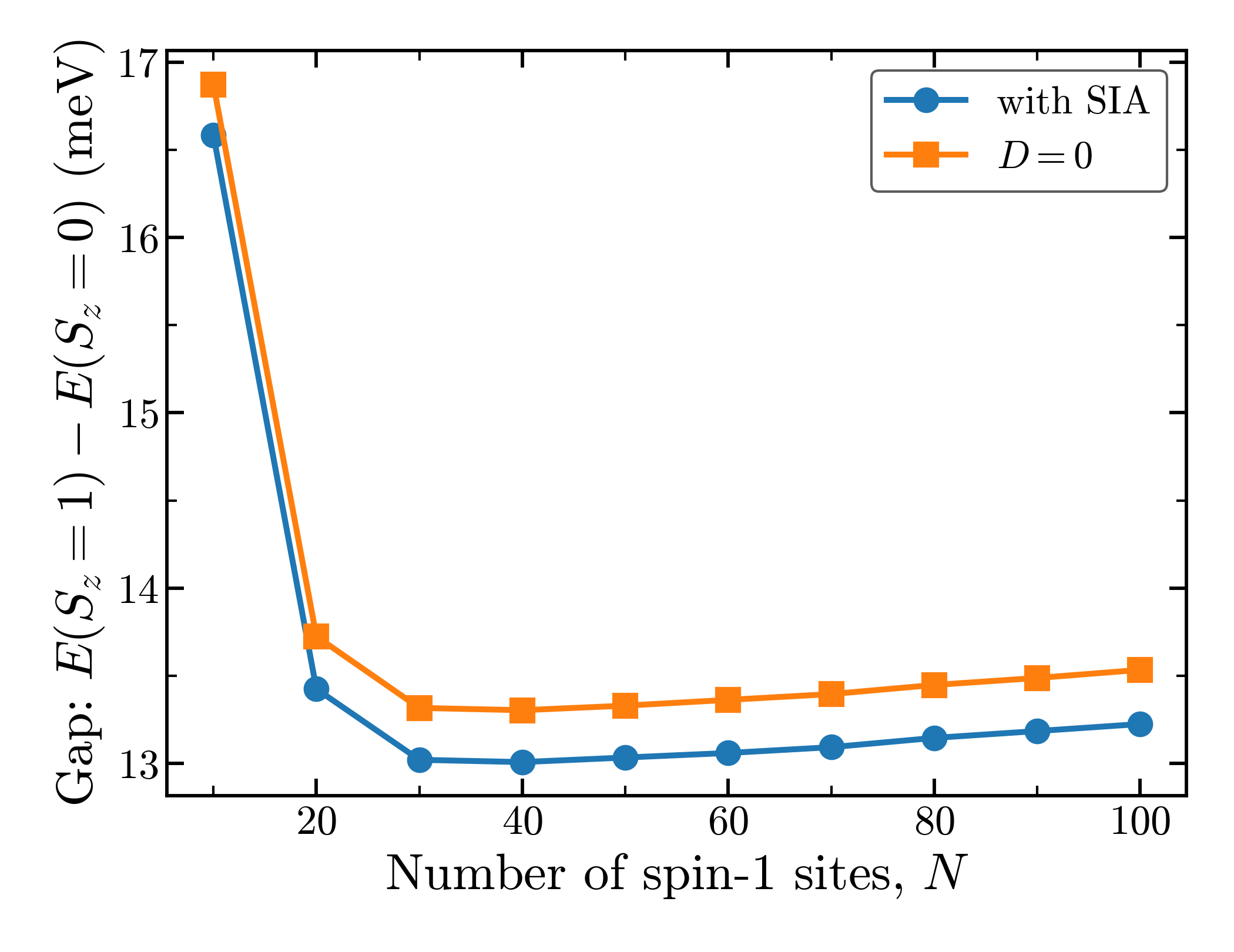}
\caption{Finite-size dependence of the lowest gap obtained from DMRG for the model with the DFT-derived diagonal SIA and for the isotropic limit. The gap changes rapidly at small $N$ and remains on the $13$--$14$~meV scale over the longer chains studied.}
\label{fig:dmrg_gap}
\end{figure}

As shown in Fig.~\ref{fig:dmrg_gap}, the anisotropic result changes from 16.613~meV at $N=10$ to 13.232~meV at $N=100$, while the corresponding isotropic values are 16.899 and 13.537~meV. The $N=100$ anisotropic value is close to the lowest ED extrapolation, 13.382~meV. We therefore use the ED and DMRG results together as evidence that the DFT-derived model remains gapped on the $13$--$14$~meV scale and that the weak SIA mainly splits the low-energy triplet-derived manifold rather than changing the overall energy scale.

These calculations establish the spectral features expected for a spin-1 chain near the isotropic Haldane limit. They do not constitute a complete demonstration of symmetry-protected topological order, which would require additional diagnostics: a nonlocal string order parameter that remains finite even where conventional two-point spin correlations decay exponentially; an entanglement spectrum whose levels are at least doubly degenerate as a consequence of the symmetries protecting the phase~\cite{Pollmann2012}; and, for an open chain, a pair of effective spin-$1/2$ degrees of freedom localized at the two ends, reflecting the fractionalization of the bulk spin-1 moments into boundary states of the kind observed directly by scanning tunnelling spectroscopy in triangulene spin chains~\cite{Mishra2021}. Evaluating these signatures for the present DFT-derived Hamiltonian is left for future work.

\section{Dynamical spin structure factor}
\label{sec:dsf}

\begin{figure*}[t]
\centering
\includegraphics[width=0.80\textwidth]{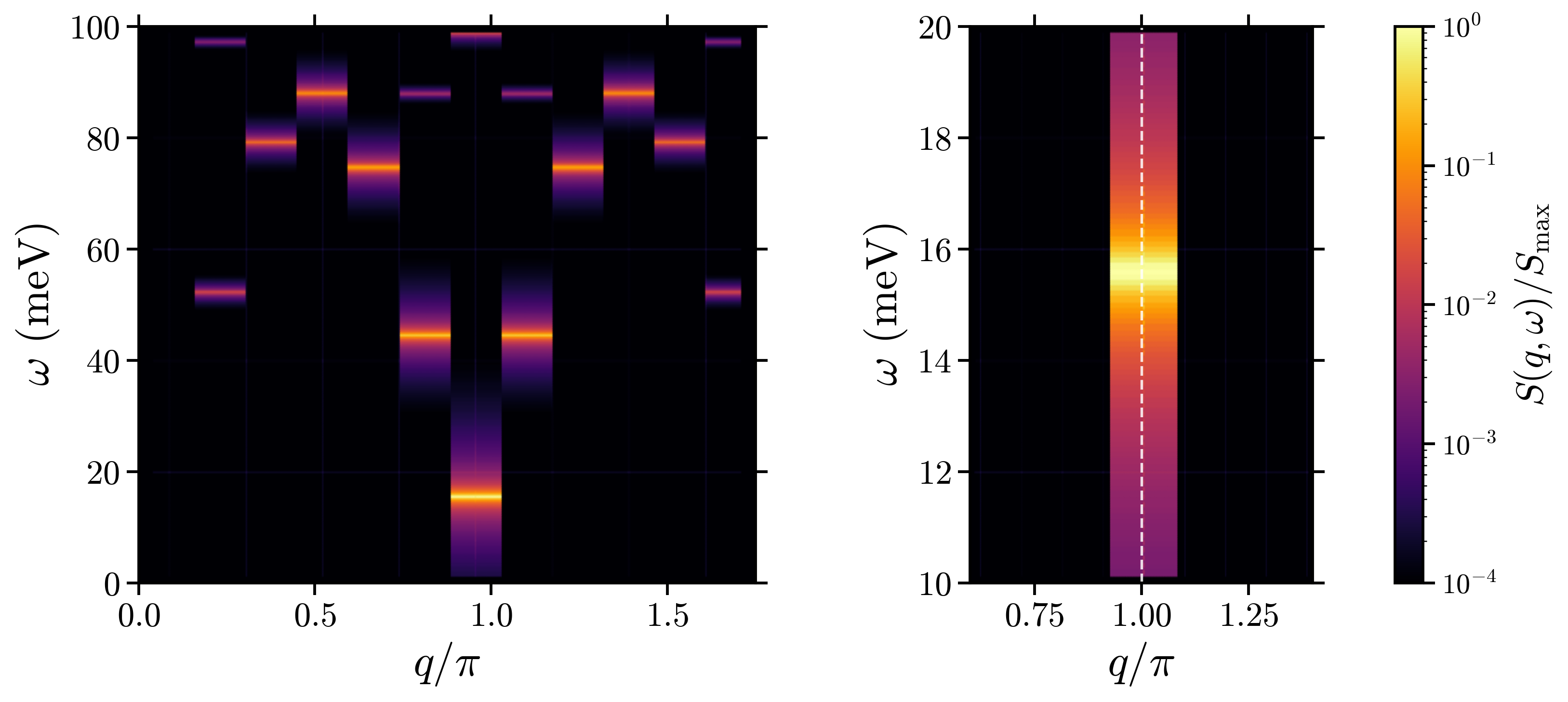}
\caption{Dynamical spin structure factor $S(q,\omega)$ for the $N=12$ periodic spin-1 chain in the isotropic limit. The left panel shows the full spectrum over 0--100~meV and the right panel enlarges the 10--20~meV range near $q=\pi$. The lowest strong response occurs at the antiferromagnetic wave vector.}
\label{fig:dsf_d0}
\end{figure*}

\begin{figure*}[t]
\centering
\includegraphics[width=0.80\textwidth]{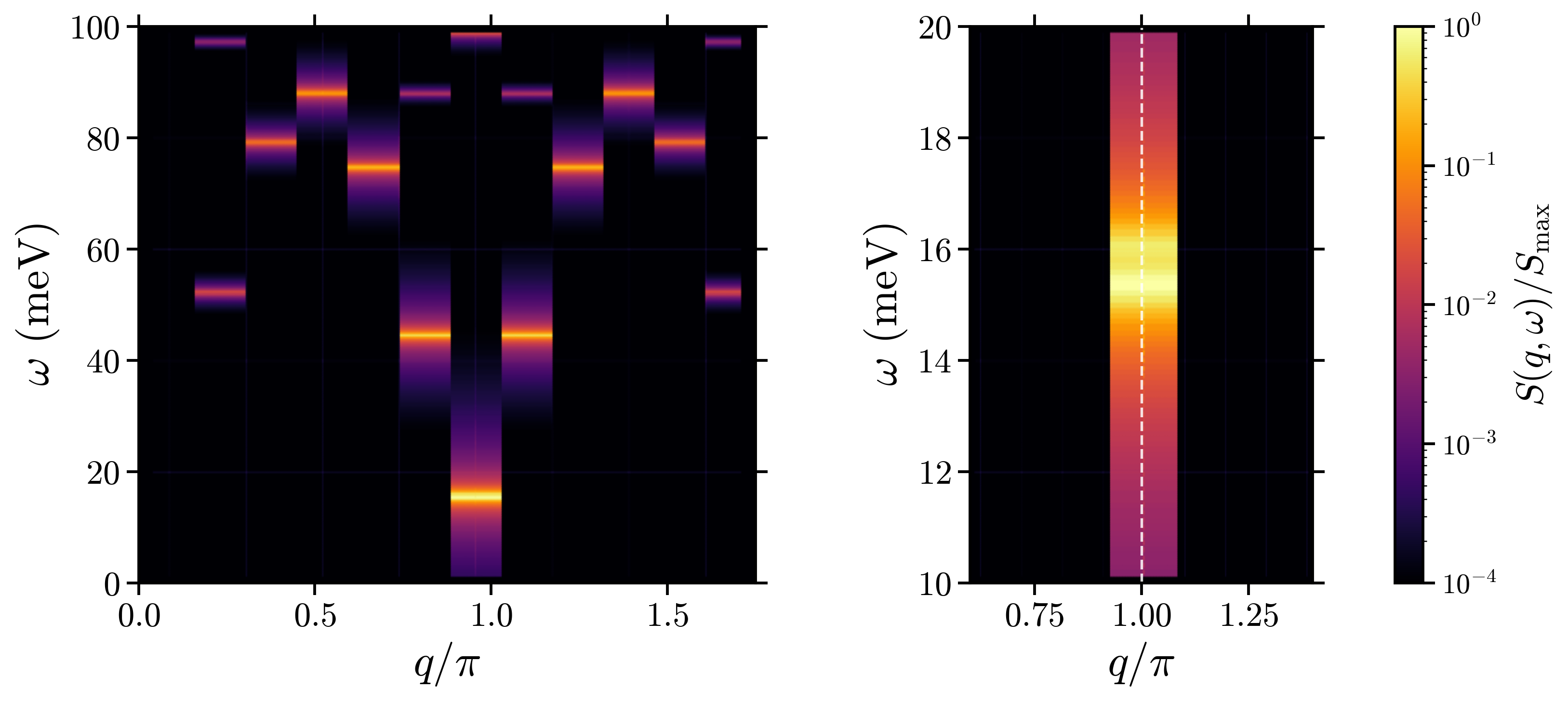}
\caption{Dynamical spin structure factor $S(q,\omega)$ for the $N=12$ periodic chain with the DFT-derived diagonal single-ion anisotropy. The same broadening, $\eta=0.25$~meV, is used as in Fig.~\ref{fig:dsf_d0}. The low-energy spectral weight remains centered at $q=\pi$, with only small changes relative to the isotropic spectrum.}
\label{fig:dsf_dft}
\end{figure*}

\begin{figure}[t]
\centering
\includegraphics[width=\columnwidth]{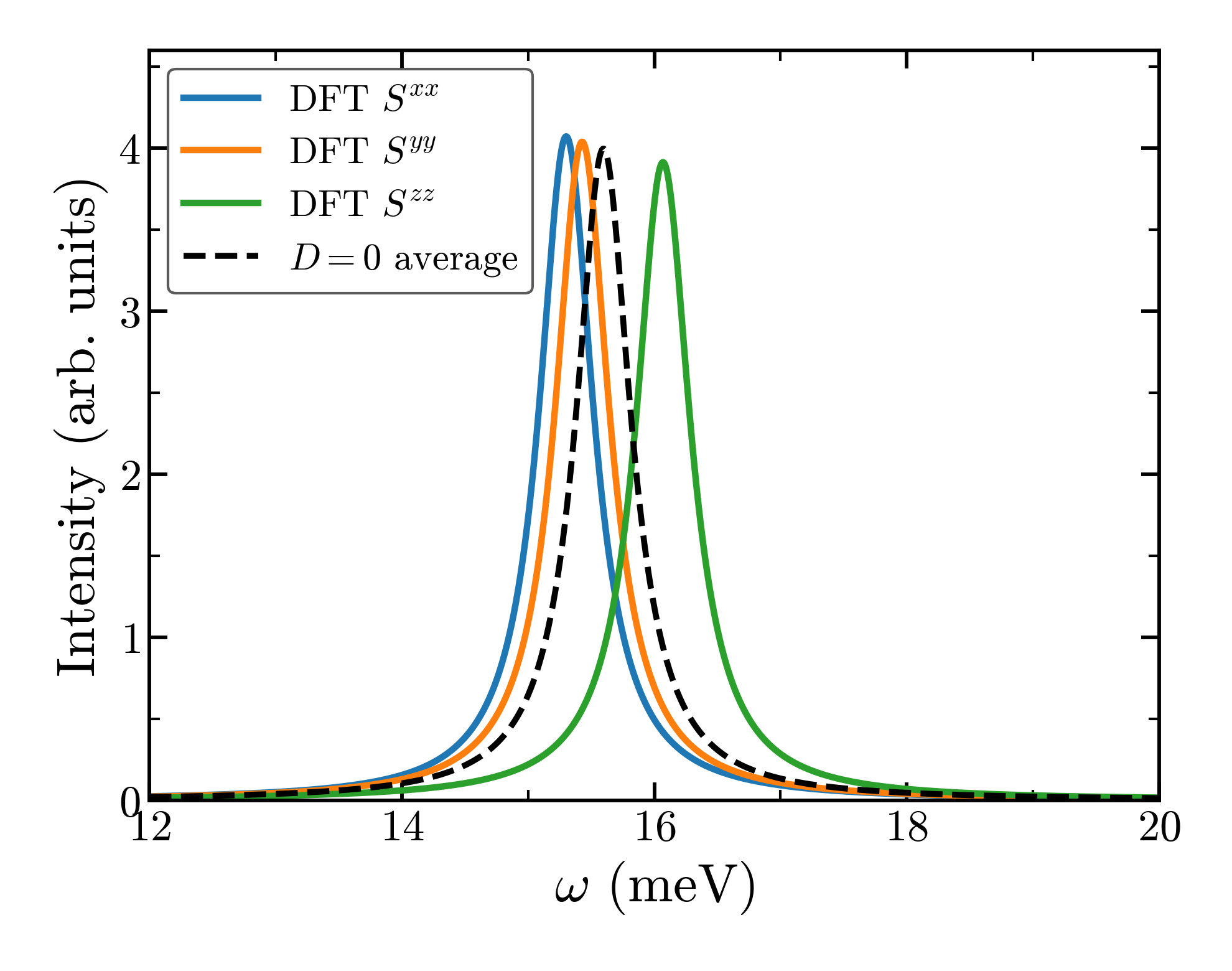}
\caption{Component-resolved $S^{\alpha\alpha}(\pi,\omega)$ for the $N=12$ periodic chain. The dashed black curve is the component-averaged isotropic response; solid curves show the anisotropic $S^{xx}$, $S^{yy}$, and $S^{zz}$ responses. The out-of-plane mode lies slightly above the in-plane modes.}
\label{fig:dsf_qpi_cut}
\end{figure}

To resolve the momentum dependence of the magnetic excitations, we calculated the zero-temperature dynamical spin structure factor for an $N=12$ periodic chain using the Lanczos continued-fraction method. The component-resolved response is
\begin{equation}
\begin{split}
S^{\alpha\alpha}(q,\omega)
=
-\frac{1}{\pi}\,\mathrm{Im}
\bigg[
\Big\langle\psi_0\Big|
\hat S_{-q}^{\alpha}
\frac{1}{\omega+E_0-\hat H+i\eta}
\hat S_q^\alpha
\Big|\psi_0\Big\rangle
\bigg],
\end{split}
\label{eq:dsf_component}
\end{equation}
with
\begin{equation}
\hat S_q^\alpha
=
\frac{1}{\sqrt N}\sum_{j=1}^N e^{iqr_j}\hat S_j^\alpha,
\end{equation}
and
\begin{equation}
S(q,\omega)
=
S^{xx}(q,\omega)+S^{yy}(q,\omega)+S^{zz}(q,\omega).
\end{equation}
A Lorentzian broadening $\eta=0.25$~meV was used. Because the chain is finite, the allowed momenta are discrete and the plotted spectra consist of broadened Lanczos poles.

Figure~\ref{fig:dsf_d0} shows the isotropic response. The lowest strong spectral feature occurs at the antiferromagnetic wave vector $q=\pi$, at approximately 15--16~meV for $N=12$, consistent with the finite-size ED gap at the same chain length. Away from $q=\pi$, the dominant spectral weight shifts to higher energy. This behavior is the finite-size precursor of the triplon dispersion expected for a Haldane chain, which near $q=\pi$ takes the approximate relativistic form $\omega(q)\simeq\sqrt{\Delta^2+v^2(q-\pi)^2}$, with $\Delta$ the Haldane gap and $v$ the spin-wave velocity of the underlying $O(3)$ nonlinear sigma model to which the low-energy theory maps~\cite{haldane1983nonlinear}. The flattening of the dispersion at $q=\pi$ and its rise away from it in Fig.~\ref{fig:dsf_d0} are qualitatively consistent with this form, although a quantitative extraction of $v$ would require chain lengths well beyond $N=12$.

The DFT-anisotropic result in Fig.~\ref{fig:dsf_dft} retains the same overall momentum dependence because the SIA is much smaller than the exchange scale. The low-energy response remains concentrated near $q=\pi$, while the anisotropy produces a small component-dependent splitting.

This splitting is clearer in the $q=\pi$ component-resolved cut in Fig.~\ref{fig:dsf_qpi_cut}. The $S^{xx}$ and $S^{yy}$ responses lie slightly below the $S^{zz}$ response, consistent with the easy-plane character of the diagonal SIA used in the many-body model.

The concentration of low-energy spectral weight at the antiferromagnetic wave vector is a spectroscopic fingerprint consistent with Haldane-chain excitations. For the present surface-derived material, the calculation should be viewed as a prediction for the intrinsic chain model; substrate coupling may modify both the energy scale and linewidths in an experimental realization. An additional representation of the momentum-dependent spectra is given in the Supplemental Material.

\section{Conclusion}
\label{sec:conclusion}

We have developed a first-principles-to-many-body description of the isolated cis-dehydroindigo coordination-polymer chain. Magnetic energy mapping gives a nearly isotropic antiferromagnetic exchange, $J_{\mathrm H}=32.206$~meV, with much weaker single-ion anisotropy, placing the spin-1 Hamiltonian close to the isotropic antiferromagnetic Heisenberg limit.

ED gives a finite low-energy gap on the 13--14~meV scale and shows that the DFT-derived diagonal SIA produces only a small splitting of the lowest triplet-derived excitation. Longer-chain DMRG calculations give the same gap scale, and the dynamical spin structure factor places the lowest strong spectral weight near $q=\pi$. Finite-chain estimates of the magnetic specific heat and susceptibility, given in the Supplemental Material, are consistent with this gapped picture but are not used here as independent evidence for it. Together, the ED, DMRG, and dynamical-structure-factor results show that the microscopic interactions of the isolated cis-dehydroindigo chain place it in a gapped regime consistent with Haldane-chain physics.

The present calculations do not evaluate string order, the entanglement spectrum, or open-chain edge states and therefore do not establish the full symmetry-protected-topological characterization. They instead provide a material-specific link between an Fe-based spin-crossover coordination polymer and the interaction regime required for Haldane-chain behavior. Explicit treatment of Ag(111) will be needed to determine how screening and hybridization renormalize this intrinsic limit.

\begin{acknowledgments}
We acknowledge the PARAM Yukti facility under the National Supercomputing Mission at the Jawaharlal Nehru Centre for Advanced Scientific Research for providing computational resources. We thank Shobhana Narasimhan (Theoretical Sciences Unit, JNCASR) for helpful discussions. We also thank Anthoula C. Papageorgiou (National and Kapodistrian University of Athens), Johannes V. Barth (Technical University of Munich), and members of their groups for the experimental realization and characterization of the Fe–indigo coordination polymers studied here, and for our longstanding collaboration on this system.

\end{acknowledgments}

\vspace{-4pt}
\bibliographystyle{apsrev4-2}
\renewcommand{\bibfont}{\footnotesize}
\bibliography{refer}

@article{haldane1983nonlinear,
  title={{Nonlinear field theory of large-spin Heisenberg antiferromagnets: semiclassically quantized solitons of the one-dimensional easy-axis N{\'e}el state}},
  author={Haldane, F Duncan M},
  journal={Phys. Rev. Lett.},
  volume={50},
  number={15},
  pages={1153},
  year={1983},
  publisher={APS},
  doi={10.1103/PhysRevLett.50.1153}
}

@article{renard1987presumption,
  title={Presumption for a quantum energy gap in the quasi-one-dimensional S= 1 Heisenberg antiferromagnet Ni (C2H8N2) 2NO2 (ClO4)},
  author={Renard, JP and Verdaguer, M and Regnault, LP and Erkelens, WAC and Rossat-Mignod, J and Stirling, WG},
  journal={EPL},
  volume={3},
  number={8},
  pages={945--952},
  year={1987},
  doi={10.1209/0295-5075/3/8/013}
}

@article{XiangWhangbo2013,
  title = {Magnetic properties and energy-mapping analysis},
  author = {Xiang, H. J. and Lee, C. and Koo, H.-J. and Gong, X. and Whangbo, M.-H.},
  journal = {Dalton Trans.},
  volume = {42},
  issue = {4},
  pages = {823--853},
  year = {2013},
  publisher = {Royal Society of Chemistry},
  doi = {10.1039/C2DT31662E}
}

@article{Udvardi2003,
  title = {First-principles relativistic study of spin waves in thin magnetic films},
  author = {Udvardi, L. and Szunyogh, L. and Palot{\'a}s, K. and Weinberger, P.},
  journal = {Phys. Rev. B},
  volume = {68},
  issue = {10},
  pages = {104436},
  numpages = {14},
  year = {2003},
  publisher = {American Physical Society},
  doi = {10.1103/PhysRevB.68.104436}
}

@article{Kresse1996,
author = {Kresse, Georg and Furthm{\"u}ller, J{\"u}rgen},
title = {Efficient iterative schemes for ab initio total-energy calculations using a plane-wave basis set},
journal = {Phys. Rev. B},
volume = {54},
number = {16},
pages = {11169--11186},
year = {1996},
doi = {10.1103/PhysRevB.54.11169}
}

@article{Kresse1999,
author = {Kresse, Georg and Joubert, D.},
title = {From ultrasoft pseudopotentials to the projector augmented-wave method},
journal = {Phys. Rev. B},
volume = {59},
number = {3},
pages = {1758--1775},
year = {1999},
doi = {10.1103/PhysRevB.59.1758}
}

@article{Blochl1994,
author = {Bl{\"o}chl, Peter E.},
title = {Projector augmented-wave method},
journal = {Phys. Rev. B},
volume = {50},
number = {24},
pages = {17953--17979},
year = {1994},
doi = {10.1103/PhysRevB.50.17953}
}

@article{Dudarev1998,
author = {Dudarev, S. L. and Botton, G. A. and Savrasov, S. Y. and Humphreys, C. J. and Sutton, A. P.},
title = {Electron-energy-loss spectra and the structural stability of nickel oxide: An LSDA+U study},
journal = {Phys. Rev. B},
volume = {57},
number = {3},
pages = {1505--1509},
year = {1998},
doi = {10.1103/PhysRevB.57.1505}
}

@article{Hobbs2000,
author = {Hobbs, D. and Kresse, G. and Hafner, J.},
title = {Fully unconstrained noncollinear magnetism within the projector augmented-wave method},
journal = {Phys. Rev. B},
volume = {62},
number = {17},
pages = {11556--11570},
year = {2000},
doi = {10.1103/PhysRevB.62.11556}
}

@article{Weinberg2017,
  author  = {Weinberg, Phillip and Bukov, Marin},
  title   = {QuSpin: a Python package for dynamics and exact diagonalisation of quantum many-body systems Part I: Spin chains},
  journal = {SciPost Phys.},
  volume  = {2},
  pages   = {003},
  year    = {2017},
  doi     = {10.21468/SciPostPhys.2.1.003}
}

@article{Weinberg2019,
  author  = {Weinberg, Phillip and Bukov, Marin},
  title   = {QuSpin: a Python package for dynamics and exact diagonalisation of quantum many-body systems Part II: Bosons, fermions and higher spins},
  journal = {SciPost Phys.},
  volume  = {7},
  pages   = {020},
  year    = {2019},
  doi     = {10.21468/SciPostPhys.7.2.020}
}

@article{hauschild2024tenpy,
  author  = {Hauschild, Johannes and others},
  title   = {Tensor Network Python ({TeNPy}) version 1},
  journal = {SciPost Phys. Codebases},
  volume  = {41},
  year    = {2024}
}

@article{Wu2013,
  title = {Suitability of chromium phthalocyanines to test Haldane's conjecture: First-principles calculations},
  author = {Wu, Wei and Harrison, N. M. and Fisher, A. J.},
  journal = {Phys. Rev. B},
  volume = {88},
  issue = {22},
  pages = {224417},
  numpages = {9},
  year = {2013},
  publisher = {American Physical Society},
  doi = {10.1103/PhysRevB.88.224417}
}

@article{Henriques2025,
  title = {Prediction of a topological phase transition in exchange alternating spin-1 nanographene chains},
  author = {Henriques, Jo{\~a}o C. G. and del Castillo, Yelko and Segundo, Ricardo and Phillips, Jan and Fern{\'a}ndez-Rossier, Joaqu{\'i}n},
  journal = {arXiv preprint},
  eprint = {2510.23555},
  archivePrefix = {arXiv},
  primaryClass = {cond-mat.str-el},
  year = {2025},
  doi = {10.48550/arXiv.2510.23555}
}

@article{Powell2026,
  title = {Competing quantum effects in spin crossover chains: Spin-orbit coupling, magnetic exchange, and elastic interactions},
  author = {Rist, F. and Nourse, H. L. and Powell, B. J.},
  journal = {Phys. Rev. B},
  volume = {113},
  issue = {3},
  pages = {035147},
  year = {2026},
  publisher = {American Physical Society},
  doi = {10.1103/zgps-3p8n}
}

@article{white1993numerical,
  title = {Numerical renormalization-group study of low-lying eigenstates of the antiferromagnetic {$S=1$} Heisenberg chain},
  author = {White, Steven R. and Huse, David A.},
  journal = {Phys. Rev. B},
  volume = {48},
  issue = {6},
  pages = {3844--3852},
  year = {1993},
  publisher = {American Physical Society},
  doi = {10.1103/PhysRevB.48.3844}
}

@article{Buyers1986,
  title = {Experimental evidence for the Haldane gap in a spin-1 nearly isotropic, antiferromagnetic chain},
  author = {Buyers, W. J. L. and Morra, R. M. and Armstrong, R. L. and Hogan, M. J. and Gerlach, P. and Hirakawa, K.},
  journal = {Phys. Rev. Lett.},
  volume = {56},
  issue = {4},
  pages = {371--374},
  year = {1986},
  publisher = {American Physical Society},
  doi = {10.1103/PhysRevLett.56.371}
}

@article{Pollmann2012,
  title = {Symmetry protection of topological phases in one-dimensional quantum spin systems},
  author = {Pollmann, Frank and Berg, Erez and Turner, Ari M. and Oshikawa, Masaki},
  journal = {Phys. Rev. B},
  volume = {85},
  issue = {7},
  pages = {075125},
  year = {2012},
  publisher = {American Physical Society},
  doi = {10.1103/PhysRevB.85.075125}
}

@article{Mishra2021,
  title = {Observation of fractional edge excitations in nanographene spin chains},
  author = {Mishra, Shantanu and Catarina, Gon{\c{c}}alo and Wu, Fupeng and Ortiz, Ricardo and Jacob, David and Eimre, Kristjan and Ma, Ji and Pignedoli, Carlo A. and Feng, Xinliang and Ruffieux, Pascal and Fern{\'a}ndez-Rossier, Joaqu{\'i}n and Fasel, Roman},
  journal = {Nature},
  volume = {598},
  pages = {287--292},
  year = {2021},
  doi = {10.1038/s41586-021-03842-3}
}

@misc{sco_paper,
  author        = {Chakraborty, Ritam and Xu, Hongxiang and Yang, Biao and Chhabra, Harshdeep Singh and Reichert, Joachim and Barth, Johannes V. and Papageorgiou, Anthoula C. and Narasimhan, Shobhana},
  title         = {Interplay between Isomerization and Spin Crossover in {1D} {Fe}-Indigo Coordination Polymers on {Ag} Substrates},
  year          = {2026},
  eprint        = {2609.03041},
  archivePrefix = {arXiv},
  primaryClass  = {cond-mat.mtrl-sci},
  doi           = {10.48550/arXiv.2609.03041}
}

@article{Anindya2026,
  author  = {Anindya, Khalid N. and Guo, Hong},
  title   = {Tunable Topological Phases in an Organic One-Dimensional Mott Chain: Exchange-Alternating {$S=1/2$} and Haldane {$S=1$}},
  journal = {ACS Nano},
  volume  = {20},
  number  = {14},
  pages   = {11209--11218},
  year    = {2026},
  doi     = {10.1021/acsnano.5c22186}
}

@article{Bera2015,
  author  = {Bera, A. K. and Lake, B. and Islam, A. T. M. N. and Janson, O. and Rosner, H. and Schneidewind, A. and Park, J. T. and Wheeler, E. and Zander, S.},
  title   = {Consequences of critical interchain couplings and anisotropy on a Haldane chain},
  journal = {Phys. Rev. B},
  volume  = {91},
  number  = {14},
  pages   = {144414},
  year    = {2015},
  doi     = {10.1103/PhysRevB.91.144414}
}

@article{hamada2014van,
  author  = {Hamada, Ikutaro},
  title   = {van der {W}aals density functional made accurate},
  journal = {Phys. Rev. B},
  volume  = {89},
  pages   = {121103},
  year    = {2014},
  doi     = {10.1103/PhysRevB.89.121103}
}

@article{lee2010higher,
  author  = {Lee, Kyuho and Murray, {\'E}amonn D. and Kong, Lingzhu and Lundqvist, Bengt I. and Langreth, David C.},
  title   = {Higher-accuracy van der {W}aals density functional},
  journal = {Phys. Rev. B},
  volume  = {82},
  pages   = {081101},
  year    = {2010},
  doi     = {10.1103/PhysRevB.82.081101}
}

@article{xu2024,
  author  = {Xu, Hongxiang and Chakraborty, Ritam and Adak, Abhishek Kumar and Das, Arpan and Yang, Biao and Meier, Dennis and Riss, Alexander and Reichert, Joachim and Narasimhan, Shobhana and Barth, Johannes V. and Papageorgiou, Anthoula C.},
  title   = {On-Surface Isomerization of Indigo within {1D} Coordination Polymers},
  journal = {Angew. Chem. Int. Ed.},
  volume  = {63},
  pages   = {e202319162},
  year    = {2024},
  doi     = {10.1002/anie.202319162}
}

\clearpage
\onecolumngrid
\begin{center}
\rule{0.9\textwidth}{0.4pt}\\[4pt]
{\large\textbf{Supplemental Material for:}}\\[2pt]
{\large\textbf{Evidence for Haldane-Chain Physics in an Fe--Dehydroindigo Coordination Polymer}}\\[6pt]
Ritam Chakraborty, Shobhana Narasimhan, and T.~V.~Ramakrishnan\\[4pt]
\rule{0.9\textwidth}{0.4pt}
\end{center}
\vspace{4pt}
\twocolumngrid

This Supplemental Material gives the bond-counting convention and magnetic configurations used in the DFT energy mapping, the cross-validation and tensor-stability analysis, additional finite-size information for the ED and DMRG calculations, an alternative representation of the dynamical spin structure factor, and finite-chain thermodynamic results.

\section{Magnetic model and bond-counting convention}
\label{si:sec:model}

Following standard magnetic energy-mapping and relativistic spin-Hamiltonian formulations~\cite{XiangWhangbo2013,Udvardi2003}, we start from the symmetric exchange-plus-single-ion-anisotropy Hamiltonian
\begin{equation}
\begin{split}
\mathcal{H}_{\mathrm{spin}}
={}&
-\sum_{\langle i,j\rangle}\sum_{\alpha,\beta}
J_{\alpha\beta}^{i,j}S_i^\alpha S_j^\beta\\
&+
\sum_i\sum_{\alpha,\beta}
D_{\alpha\beta}^{i}S_i^\alpha S_i^\beta,
\end{split}
\label{eq:gen_hamiltonian}
\end{equation}
where $\alpha,\beta\in\{x,y,z\}$. The chain direction is taken as $x$. The two Fe sites in the magnetic unit cell have equivalent nearest-neighbor bonds, one within the cell and one connecting neighboring cells. With equal nearest-neighbor tensors on these two bonds, the periodic two-site-cell energy becomes
\begin{equation}
\mathcal{H}_{\mathrm{spin}}
=
-2\sum_{\alpha,\beta}J_{\alpha\beta}S_1^\alpha S_2^\beta
+
\sum_{\alpha,\beta}D_{\alpha\beta}
\left(S_1^\alpha S_1^\beta+S_2^\alpha S_2^\beta\right).
\label{eq:two_site_hamiltonian}
\end{equation}
This is the origin of the factor of two in the exchange convention used in the main text. Consequently, the fitted isotropic value $J_{\mathrm{iso}}=-16.103$~meV corresponds to the conventional Heisenberg coupling $J_{\mathrm H}=-2J_{\mathrm{iso}}=32.206$~meV.

The spin-only mapping also assumes that the low-energy states can be represented by essentially fixed local $S=1$ moments. In the present GGA+$U$ description, $U_{\mathrm{eff}}=1$~eV and the characteristic hopping estimate $t\simeq40$~meV correspond to $U_{\mathrm{eff}}/t\simeq25$, providing a clear separation between charge motion and the magnetic energy scale. This is the same strong-coupling logic used for organic Mott/Haldane chains, where $U/t>100$ was found to justify a spin-only Heisenberg description~\cite{Anindya2026}, and is consistent with the established GGA+$U$--to--Heisenberg mapping for the inorganic $S=1$ Haldane compound SrNi$_2$V$_2$O$_8$~\cite{Bera2015}. Since the Fe--dehydroindigo manifold is multiorbital and ligand mediated, the ratio quoted here is used as a scale-separation estimate rather than as a literal one-band Hubbard parametrization; the exchange and anisotropy tensors are determined directly from the DFT total-energy mapping.

The DFT energy for each magnetic configuration is written as
\begin{equation}
E_{\mathrm{DFT}}
=
E_0+\langle\mathcal{H}_{\mathrm{spin}}\rangle,
\end{equation}
where $E_0$ is a spin-independent reference energy. Both $\mathbf J$ and $\mathbf D$ are taken to be symmetric. The trace of $\mathbf D$ is fixed to zero because an isotropic contribution gives only a constant $S(S+1)$ shift. The fit therefore contains six independent exchange components, five independent SIA components, and the reference energy $E_0$.

For a unit vector $\mathbf n$ and collinear FM/AFM pairs, $\mathbf S_1=\mathbf n$ and $\mathbf S_2=\pm\mathbf n$, the magnetic contributions are
\begin{align}
E_{\mathrm{FM}}(\mathbf n)-E_0
&=
-2\,\mathbf n^{T}\mathbf J\,\mathbf n
+
2\,\mathbf n^{T}\mathbf D\,\mathbf n,\\
E_{\mathrm{AFM}}(\mathbf n)-E_0
&=
+2\,\mathbf n^{T}\mathbf J\,\mathbf n
+
2\,\mathbf n^{T}\mathbf D\,\mathbf n.
\end{align}
Thus the FM--AFM difference isolates the exchange contribution, while their sum constrains the SIA contribution. Sampling several off-axis magnetization directions provides the information required to determine the symmetric off-diagonal tensor elements.

\section{Magnetic configurations used for energy mapping}
\label{si:sec:configs}

The dataset contains 21 magnetization directions, each evaluated in both FM and AFM alignment, for a total of 42 DFT energies. Fifteen high-symmetry directions were used as the primary fitting set and six additional off-axis directions were used as an out-of-sample validation set. Grouped leave-one-direction-out refits were additionally used to assess the stability of the extracted tensors.

\newpage
\subsection{High-symmetry directions}

\begin{table}[h!]
\centering
\renewcommand{\arraystretch}{1.0}
\setlength{\tabcolsep}{3pt}
\footnotesize
\begin{tabular}{ccc}
\toprule
\textbf{Primary axes} & \textbf{Face diagonals} & \textbf{Body diagonals}\\
$(S_x,S_y,S_z)$ & $(S_x,S_y,S_z)$ & $(S_x,S_y,S_z)$\\
\midrule
$[1.00,0.00,0.00]$ & $[0.71,0.71,0.00]$ & $[0.58,0.58,0.58]$\\
$[0.00,1.00,0.00]$ & $[0.71,0.00,0.71]$ & $[0.58,-0.58,0.58]$\\
$[0.00,0.00,1.00]$ & $[0.00,0.71,0.71]$ & $[0.58,0.58,-0.58]$\\
 & $[0.71,-0.71,0.00]$ & $[0.58,-0.58,-0.58]$\\
 & $[0.71,0.00,-0.71]$ & $[-0.58,0.58,-0.58]$\\
 & $[0.00,0.71,-0.71]$ & $[-0.58,-0.58,0.58]$\\
\bottomrule
\end{tabular}
\caption{Fifteen high-symmetry magnetization directions used in the primary fit. Values are shown to two decimal places; $1/\sqrt{2}=0.707$ and $1/\sqrt{3}=0.577$. Each direction was calculated in both FM and AFM alignment.}
\label{tab:training_set}
\end{table}

\subsection{Off-axis validation directions}

\begin{table}[h!]
\centering
\renewcommand{\arraystretch}{1.0}
\footnotesize
\begin{tabular}{c}
\toprule
\textbf{Validation vectors} $(S_x,S_y,S_z)$\\
\midrule
$[0.230,\,0.710,\,0.660]$\\
$[0.580,\,0.170,\,0.800]$\\
$[0.410,\,0.630,\,0.650]$\\
$[0.120,\,0.910,\,0.390]$\\
$[0.770,\,0.310,\,0.560]$\\
$[0.490,\,0.440,\,0.750]$\\
\bottomrule
\end{tabular}
\caption{Six additional off-axis magnetization directions used for out-of-sample validation. Each direction was calculated in both FM and AFM alignment.}
\label{tab:testing_set}
\end{table}

\section{Cross-validation and tensor stability}
\label{si:sec:lodocv}

The FM and AFM configurations associated with a given magnetization direction were treated as a group during validation. This avoids retaining one member of a paired direction while testing the other. In the grouped leave-one-direction-out procedure, an entire direction was removed and the remaining configurations were refitted.

The reported aggregate validation metrics are
\begin{itemize}
\item $R^2=0.9999$,
\item mean absolute error (MAE) $=0.130$~meV,
\item root-mean-square error (RMSE) $=0.166$~meV,
\end{itemize}
with a maximum absolute residual below 0.38~meV. These errors are small compared with both the exchange scale and the low-energy gap scale. The validation therefore supports the use of the bilinear exchange-plus-SIA model for the sampled magnetic configurations. It should not be interpreted as a proof that all possible higher-order or longer-range interactions vanish.

\subsection{Stability of the fitted tensors}
\label{si:sec:stability}

Across the grouped validation folds, the fitted tensor components vary only weakly. The mean values, standard deviations, and largest deviations from the mean are listed below.

\noindent\textbf{Exchange tensor components:}
\begin{itemize}
\item $J_{xx}=-16.001\pm0.013$~meV \quad ($\Delta_{\max}=0.036$~meV),
\item $J_{yy}=-16.124\pm0.010$~meV \quad ($\Delta_{\max}=0.025$~meV),
\item $J_{zz}=-16.185\pm0.011$~meV \quad ($\Delta_{\max}=0.021$~meV),
\item $J_{xy}=-0.024\pm0.006$~meV \quad ($\Delta_{\max}=0.015$~meV),
\item $J_{xz}=+0.020\pm0.012$~meV \quad ($\Delta_{\max}=0.034$~meV),
\item $J_{yz}=-0.021\pm0.009$~meV \quad ($\Delta_{\max}=0.026$~meV).
\end{itemize}

\noindent\textbf{Single-ion anisotropy components:}
\begin{itemize}
\item $D_{xx}=-0.152\pm0.005$~meV \quad ($\Delta_{\max}=0.011$~meV),
\item $D_{yy}=-0.089\pm0.004$~meV \quad ($\Delta_{\max}=0.010$~meV),
\item $D_{zz}=+0.241\pm0.005$~meV \quad ($\Delta_{\max}=0.012$~meV),
\item $D_{xy}=-0.006\pm0.002$~meV \quad ($\Delta_{\max}=0.004$~meV),
\item $D_{xz}=+0.055\pm0.006$~meV \quad ($\Delta_{\max}=0.017$~meV),
\item $D_{yz}=+0.040\pm0.006$~meV \quad ($\Delta_{\max}=0.015$~meV).
\end{itemize}

The observed variations, $\Delta J_{\max}<0.04$~meV and $\Delta D_{\max}<0.02$~meV, are small on the scale of the dominant exchange interaction. This supports the numerical stability of the fitted parameters within the chosen model.

\section{Finite-size sensitivity of the exact-diagonalization gap}
\label{si:sec:ed_fits}

The main text uses an exponential finite-size form. This choice is consistent with the gapped spin-1 Heisenberg chain, whose correlations decay exponentially in the thermodynamic limit~\cite{white1993numerical}:
\begin{equation}
\Delta_a(N)
=
\Delta_a(\infty)+A_a\exp(-N/\lambda_a),
\end{equation}
which is consistent with the exponentially decaying finite-size corrections expected in a gapped one-dimensional system. The fitted decay parameters for the three anisotropy-split branches are
\begin{equation}
\lambda_1=3.666,\qquad
\lambda_2=3.662,\qquad
\lambda_3=3.645.
\end{equation}
These parameters describe the decay of the finite-size gap corrections and are not identified with the bulk spin--spin correlation length.

As a fit-form sensitivity check, the same ED data were also fitted to
\begin{equation}
\Delta_a(N)
=
\Delta_a(\infty)+\frac{b_a}{N}+\frac{c_a}{N^2}.
\end{equation}
This empirical form gives extrapolated values of 10.70, 10.85, and 11.60~meV for the three anisotropic branches and 11.05~meV in the isotropic case. The spread between the two extrapolation forms reflects the limited system sizes accessible to ED. For this reason, the main text does not use the inverse-size fit as an independent estimate of the thermodynamic Haldane gap; instead, it uses the exponential fit together with the longer-chain DMRG results and the known isotropic scale as complementary information.

\section{Extended DMRG--ED comparison}
\label{si:sec:dmrg}

\begin{figure}[t]
\centering
\includegraphics[width=0.96\columnwidth]{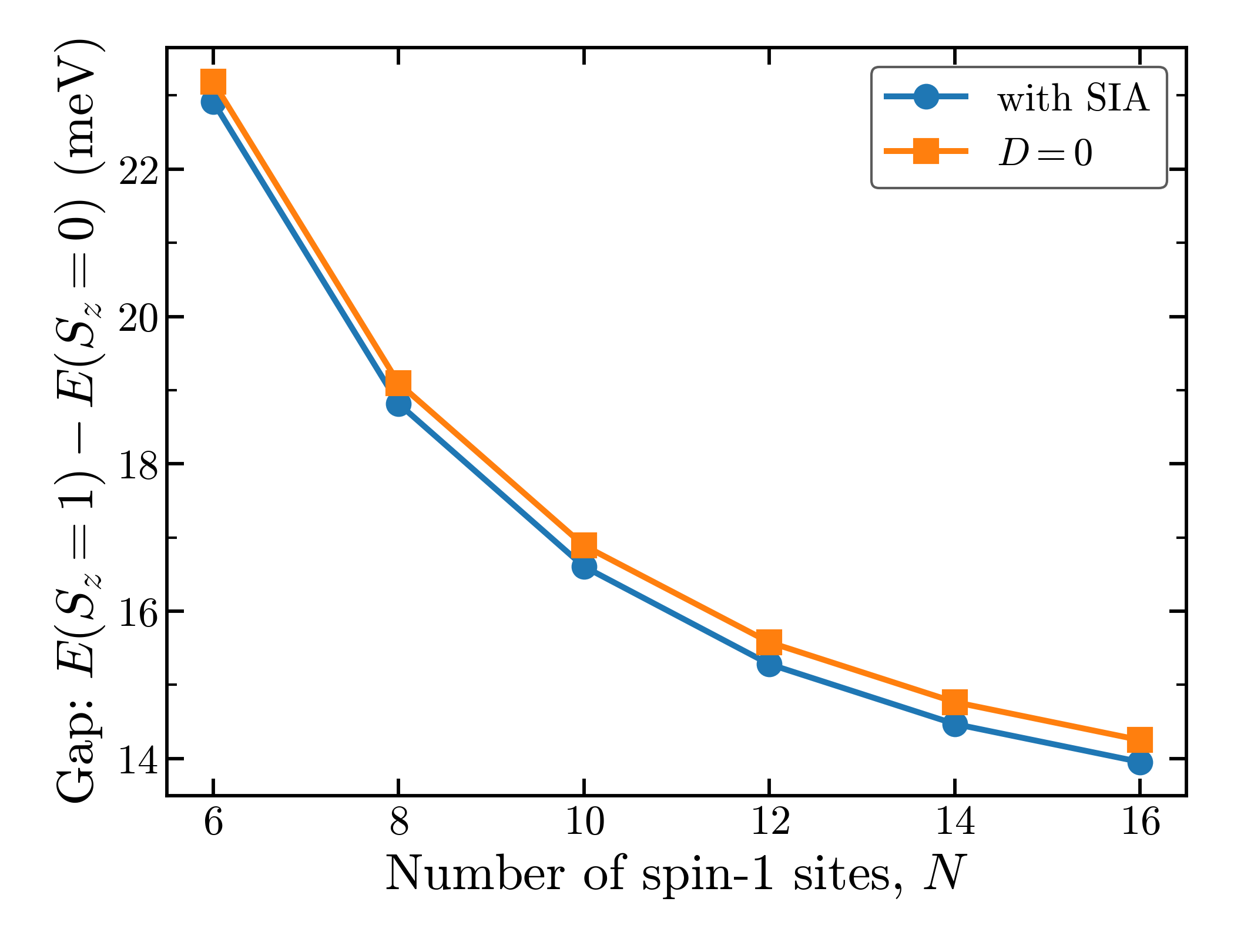}
\caption{Finite-size dependence of the lowest DMRG gap for the model with the DFT-derived diagonal SIA and for the isotropic limit, evaluated over the same chain lengths, $N=6$--16, used for the exact-diagonalization gap in the main text. In the present implementation the plotted quantity is $\Delta_{\mathrm{DMRG}}(N)=E_0(S^z_{\mathrm{tot}}=1)-E_0(S^z_{\mathrm{tot}}=0)$. The figure provides a direct DMRG-ED comparison at matching chain lengths.}
\label{fig:DMRG_ED_comparison}
\end{figure}

Figure~\ref{fig:DMRG_ED_comparison} shows the DMRG gap evaluated over the same range of chain lengths, $N=6$--16, used for the ED calculation in the main text, giving a direct method-to-method comparison at matching $N$. The DMRG gap tracks the lowest ED branch closely over this range. As noted in the main text, the anisotropic DMRG calculation conserves $S^z_{\mathrm{tot}}$ by retaining only the $U(1)$-symmetric part of the single-ion anisotropy, $D_\perp=(D_{xx}+D_{yy})/2$ together with $D_{zz}$; the omitted symmetry-breaking term, $(D_{xx}-D_{yy})/2\simeq0.032$~meV, is roughly three orders of magnitude smaller than $J_{\mathrm H}$ and two orders smaller than $D_{zz}$ itself, and is not expected to shift the reported gap values within their quoted precision. The full orthorhombic anisotropy is retained without approximation in the ED calculation shown alongside it.

\section{Additional dynamical-structure-factor representation}
\label{si:sec:dsf}

\begin{figure}[t]
\centering
\includegraphics[width=0.96\columnwidth]{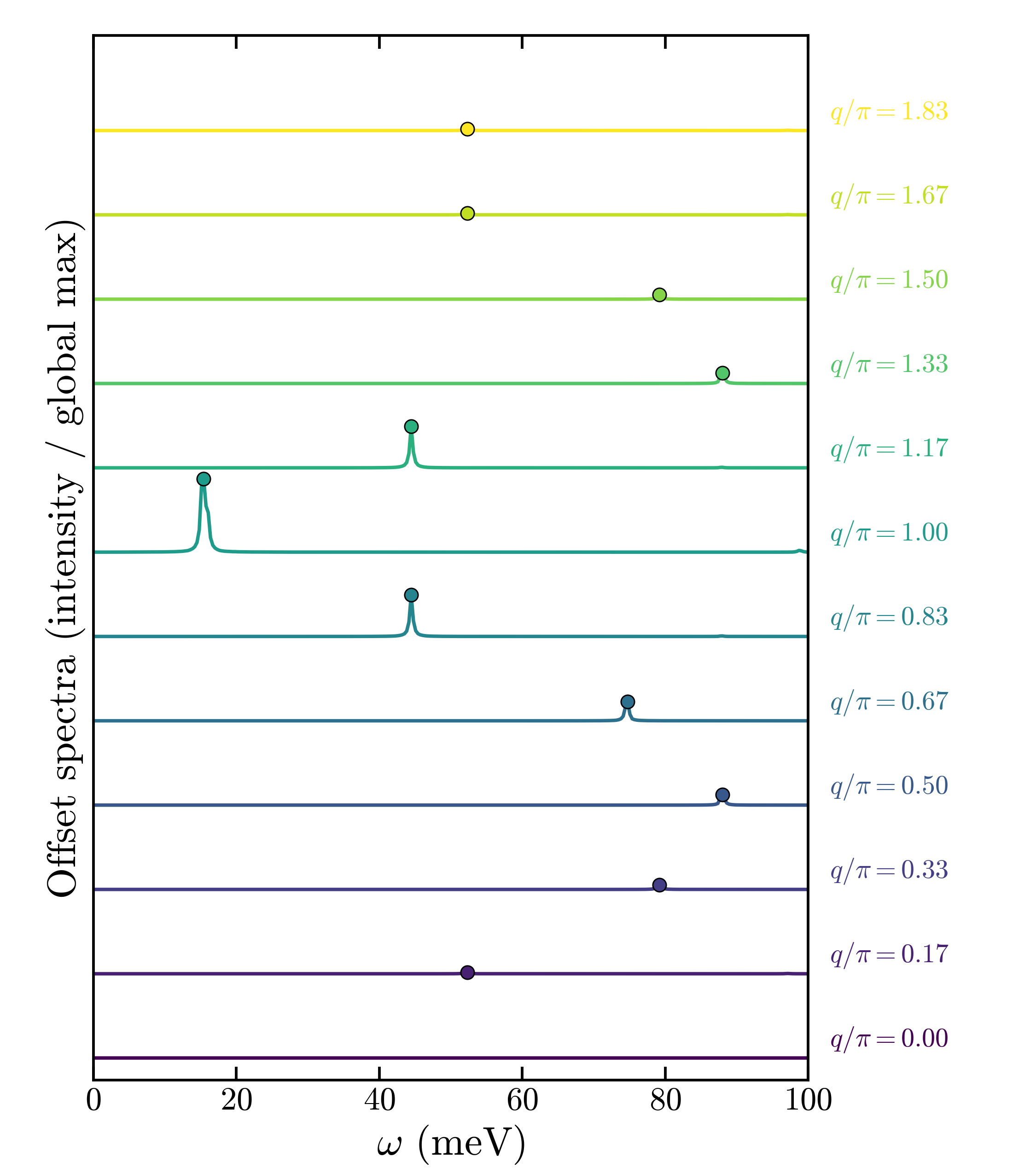}
\caption{Stacked frequency cuts of the DFT-anisotropic dynamical spin structure factor for the $N=12$ periodic chain. Each curve corresponds to one allowed momentum, and the marked points indicate the dominant peak.}
\label{fig:dsf_stacked_si}
\end{figure}

Figure~\ref{fig:dsf_stacked_si} shows stacked frequency cuts of the DFT-anisotropic $S(q,\omega)$ for the discrete momenta of the $N=12$ periodic chain. The dominant low-energy feature occurs at $q=\pi$, while the peak position shifts to higher energy away from the antiferromagnetic wave vector. This momentum dependence traces out the finite-size approximation to the triplon dispersion discussed in the main text; the systematic rise of the peak energy away from $q=\pi$ is the discrete-momentum signature of the dispersive branch $\omega(q)\simeq\sqrt{\Delta^2+v^2(q-\pi)^2}$ expected on general grounds for a gapped one-dimensional antiferromagnet, with the curvature near $q=\pi$ controlled by the spin-wave velocity $v$.

\section{Finite-chain thermodynamic estimates}
\label{si:sec:thermo}

For completeness, we evaluated thermodynamic quantities from the complete spectrum of an $N=8$ periodic spin-1 chain. These calculations are finite-size model estimates and are not used to establish the Haldane-like character of the chain in the main text. In particular, they do not include temperature-dependent changes of the spin state, coupling to Ag(111), or other structural degrees of freedom that may become relevant in the experimental material.

For temperature $T$, the partition function is
\begin{equation}
Z
=
\sum_n\exp[-\beta(E_n-E_0)],
\qquad
\beta=(k_BT)^{-1}.
\end{equation}
The magnetic contribution to the specific heat is calculated from the energy fluctuations,
\begin{equation}
C_{\mathrm{mag}}(T)
=
\frac{N_A}{N}
\frac{\langle E^2\rangle-\langle E\rangle^2}{k_BT^2}.
\label{eq:specific_heat}
\end{equation}

The calculated $C_{\mathrm{mag}}(T)$ is suppressed at low temperature and develops a broad maximum near 300--350~K. This is consistent with the exchange scale $J_{\mathrm H}=32.206$~meV, corresponding to approximately 374~K. The weak SIA produces little visible change on this scale.

For the isotropic model, the zero-field molar susceptibility can be written as the magnetization-fluctuation expression
\begin{equation}
\chi_{\mathrm m}(T)
=
\frac{\mu_0N_A(g\mu_B)^2}{Nk_BT}
\left[
\left\langle\left(\sum_iS_i^z\right)^2\right\rangle
-
\left\langle\sum_iS_i^z\right\rangle^2
\right],
\label{eq:susceptibility}
\end{equation}
with $g\simeq2$. The same fluctuation estimator was evaluated for the weakly anisotropic finite-chain model to show the small numerical effect of the SIA in the present data set.

\begin{figure}[t]
\centering
\includegraphics[width=0.96\columnwidth]{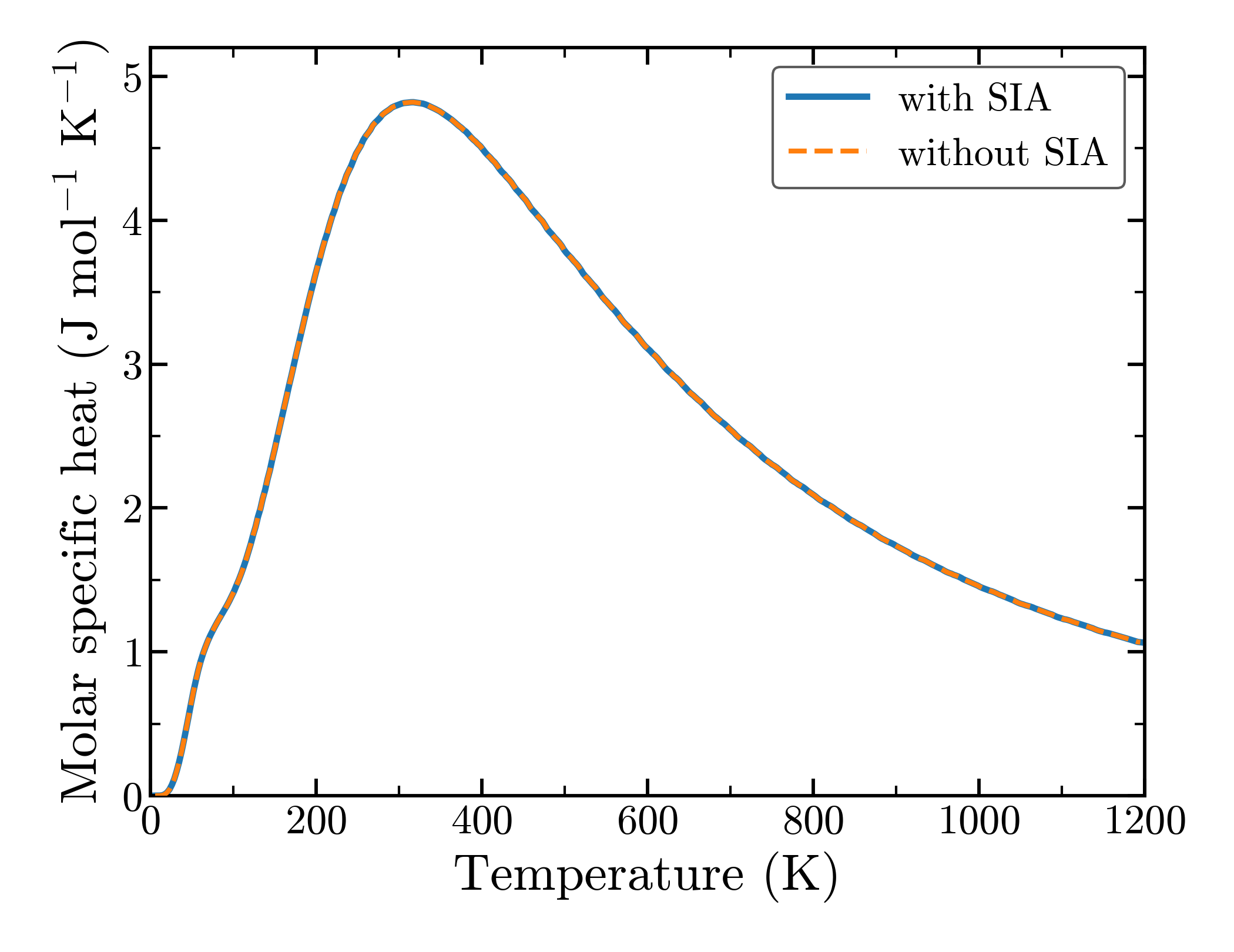}
\caption{Finite-chain magnetic specific heat for the $N=8$ periodic spin-1 model, with and without the DFT-derived diagonal SIA. The broad maximum occurs on the exchange-energy scale, while the two curves are nearly indistinguishable because the anisotropy is weak compared with $J_{\mathrm H}$.}
\label{fig:specific_heat_si}
\end{figure}

\begin{figure}[t]
\centering
\includegraphics[width=0.96\columnwidth]{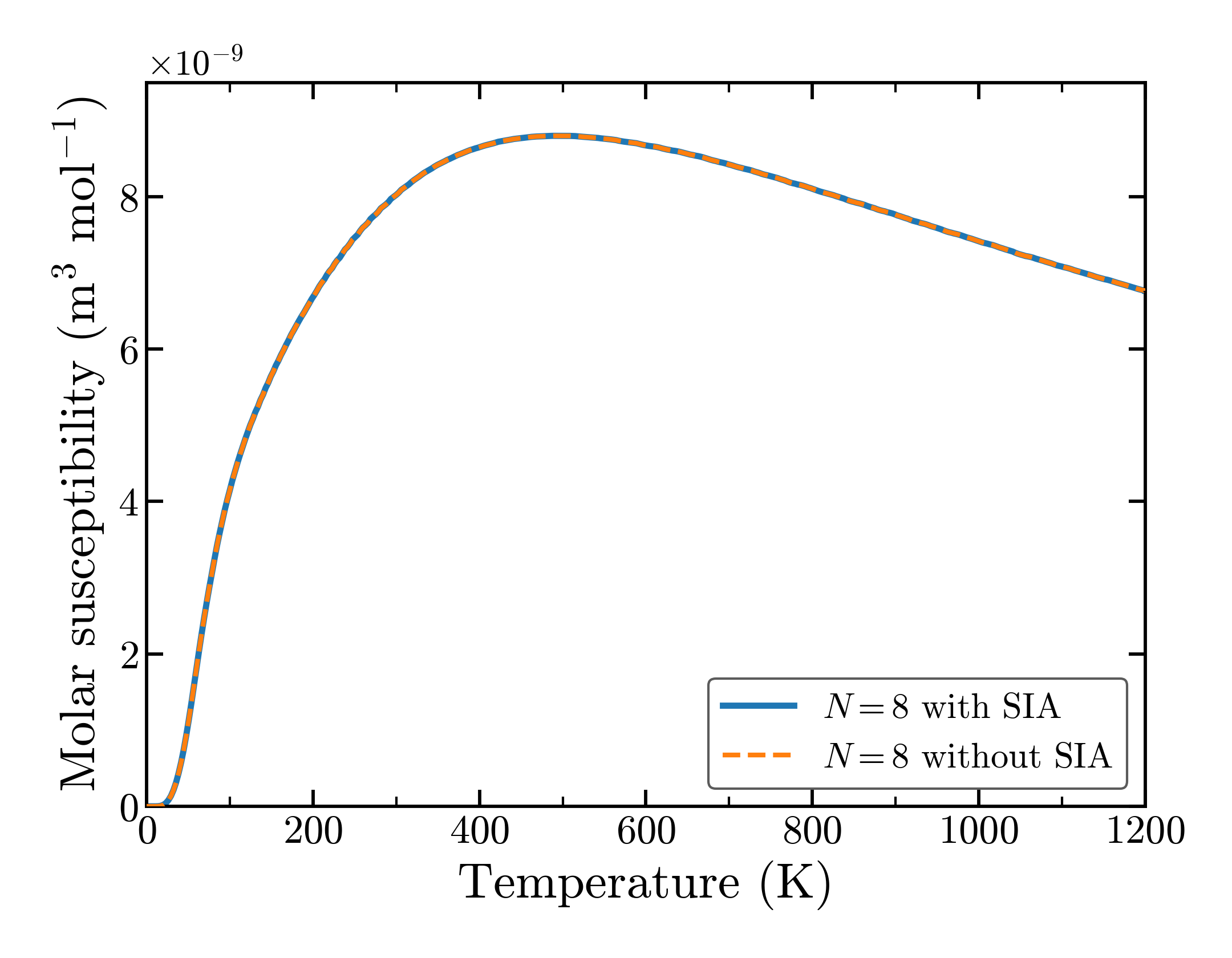}
\caption{Magnetization-fluctuation estimate for the $N=8$ periodic chain. The response is suppressed at low temperature and develops a broad maximum characteristic of short-range antiferromagnetic correlations. The DFT-derived diagonal SIA has little effect on the scale of the plot.}
\label{fig:susceptibility_si}
\end{figure}

The broad thermodynamic features are consistent with a finite-size antiferromagnetic spin chain whose dominant energy scale is set by $J_{\mathrm H}$. The absence of a sharp anomaly, in place of a Schottky-like maximum, is itself characteristic of a one-dimensional magnet: short-range antiferromagnetic correlations build up gradually along the chain as $T$ decreases, without the long-range order that would produce a true finite-temperature phase transition. At temperatures well below the gap, $k_BT\ll\Delta$, both $C_{\mathrm{mag}}(T)$ and $\chi_{\mathrm m}(T)$ are expected to be thermally activated, $\propto\exp(-\Delta/k_BT)$, reflecting the energy cost of populating the lowest triplet-derived excitation above the singlet ground state at $\Delta\approx13$--$14$~meV (Sec.~\ref{sec:manybody}); on the $N=8$ chain studied here this activated regime is only partially resolved before finite-size effects set in, so we do not attempt to extract $\Delta$ from these curves independently. Because $N=8$ is small and the material is a spin-crossover coordination polymer, these curves are best regarded as supporting model calculations rather than quantitative predictions for the full temperature dependence of the surface-supported system.

\end{document}